\documentclass[11pt]{article}
\usepackage{graphicx} 
\usepackage{bm}
\usepackage{amsmath}%
\usepackage{amsfonts}
\usepackage{amssymb}
\usepackage{tikz}
\usepackage{mathtools}
\usepackage{caption}
\usepackage[margin=1in]{geometry}

\usepackage[round,authoryear]{natbib}
\setcitestyle{aysep={,}}

\usepackage{url}

\usepackage{bm}

\title{The reflection-transmission problem for inertial waves on geostrophic shear layers}

\author{Lennart Kira$^{* \, 1}$, Jerome Noir$^1$ and Daniel Lecoanet$^2$}
\date{September 2025}

\begin{document}
\maketitle

* Corresponding author:  \textcolor{blue}{lennart.kira@eaps.ethz.ch}\\
1 Dept. of Earth and Planetary Sciences, ETH Zurich, Sonneggstrasse 5, 8092 Zurich, CH\\
2 Dept. of Engineering Sciences and Applied Mathematics, Northwestern University, 2145 Sheridan Road, Evanston, IL 60208  

\section*{Abstract}
Inertial waves in fluid regions of planets and stars play an important role in their dynamics and evolution, through energy, heat and angular momentum transport and mixing of chemicals.
While inertial wave propagation in flows prescribed by solid-body rotation is well-understood, natural environments are often characterized by convection or zonal flows.
In these more realistic configurations, we do not yet understand the propagation of inertial waves or their transport properties.
In this work, we focus on the interaction between inertial waves and geostrophic currents, which has thus far only been investigated using ray theory, where the wave length is assumed to be small relative to the length scale of the current, or averaging/statistical approaches.
We develop a quasi-two-dimensional analytical model to investigate the reflection and transmission of inertial waves in the presence of a localized geostrophic shear layer of arbitrary width and compare our theoretical findings to a set of numerical simulations.
We demonstrate that, in contrast to ray theory predictions, partial reflections occur even in subcritical shear layers and tunnelling with almost total transmission is possible in supercritical shear layers, if the layer is thin compared to the wavelength.
That is, supercritical shear layers act as low-pass filters for inertial wave beams allowing the low-wavenumber waves to travel through.
Thus, our analytical model allows us to predict interactions between inertial waves and geostrophic shear layers not addressed by ray-based or statistical theories and conceptually understand the behaviour of the full wavefield around and inside such layers.


\section{Introduction}
\label{sec:Introduction}
Inertial waves are propagating features inside rotating fluids, which are restored by the Coriolis force. They are widely observed in Earth's oceans and represent a significant peak in the kinetic energy spectrum \citep{gonella1972rotary, munk1981internal}. 
In liquid planetary layers, inertial waves can be excited by mechanical forcings and they are thought to be important for orbital and spin evolution of celestial bodies (for a review, see \citealt{le2015flows}).
Furthermore, inertial modes have been observed in the sun \citep{hanson2022discovery, triana2022identification} and their discovery has sparked questions about whether these can be used to infer interior solar properties to which acoustic modes---as canonically utilized in helioseismology---are insensitive.
Inertial modes have also been detected in $\gamma$ Doradus stars, where they can be used to probe the rotation rate of the convective cores of these stars \citep{Ouazzani2020,Saio2021}.
In most geophysical systems, stratification of the fluid introduces gravity as a second restoring force, generating inertia-gravity waves. Wave-like features in the atmospheres of gas planets are frequently interpreted as such inertia-gravity waves (\citealt{orton2020survey, rogers2016dispersive} and references therein).\\
As all of the examples listed above also include background flows in addition to waves, it is important to understand the interaction between inertial waves, moving on fast time scales, and slowly varying background flows.
The effect of a background flow shear on inertia-gravity waves has been studied by researchers in oceanography. \citet{kunze1985near} describes the propagation of near-inertial waves through geostrophic currents using ray-theory.
His calculations show a geostrophic shear can reduce the background vorticity to prohibit small scale near-inertial waves from travelling. On the other hand, he finds that waves excited in regions where the geostrophic shear increases the background vorticity may not be able to exit the shear zone and get trapped inside a jet. A similar problem has been studied by \citet{baruteau2012inertial}, who consider inertial wave attractors in spherical shells affected by differential rotation.
However, ray theory assumes that the spatial scales of the background shear are large compared to the wavelength of the propagating inertial oscillations. The method cannot describe the propagation of inertial waves inside geostrophic flows on length scales of or below their wavelengths.\\
Another approach is to use scattering theory to model the interaction between small scale geostrophic flows---which are treated analogously as a scattering potential in quantum mechanics---and near-inertial wave fields \citep{olbers1981formal}. The author considers an example of inertia-gravity waves horizontally entering ocean fronts at different angles, where the fronts are characterized by spatially varying buoyancy fields and geostrophic flows of different widths. Although he finds that waves entering along front generate a stronger scattered far-field, insights on the fundamental relations between the scattered field, the length scale of the jets and the characteristics of the incident waves are not examined. Furthermore, the scattering theory only provides insight into the dynamics of the far-field away from the sheared region and not close to or inside of it.\\
\citet{young1997propagation} study the effect of small-scale, geostrophic eddies on near-inertial modes. They find that geostrophic background turbulence introduces a frequency shift in high-wavenumber vertical modes which are thus dispersed vertically. As a consequence, near-inertial oscillations excited at the ocean's surface are efficiently transported hundreds of meters in depth within a few weeks time.
Relying on an averaging procedure for the wave field, they find that 
waves are only affected by the geostrophic flow averaged on the length scale of the waves.\\
While averaging and scattering methods, such as the ones of \citet{olbers1981formal} or \citet{young1997propagation}, successfully explain observations of anomalous near-inertial wave propagation due to smaller scale geostrophic jets or vortices, the fundamental mechanisms of inertial waves interacting with small-scale shear zones are poorly understood. However, from all previous studies, it is apparent that background currents can significantly alter the behaviour of inertial waves and a thorough understanding of the physics involved is crucial for us to comprehend the role of these waves in dynamical, planetary fluid regions.\\
In this paper, we study inertial wave propagation in a simple quasi-two-dimensional geostrophic shear layer. For this problem, we first present a theory describing the reflection-transmission behaviour of inertial plane waves entering a discrete layer of constant geostrophic shear and arbitrary width. This serves to develop conceptual understanding of the interaction between inertial waves and a shear layer.
This calculation is analogous to previous analyses of internal gravity waves interacting with a layer with different buoyancy frequency \citep{sutherland2004internal}.
We then leverage the results for a discrete layer to find the transmission and reflection behaviour of inertial waves interacting with continuous shear layers of arbitrary horizontal structure. As for the case of internal gravity waves in layered fluids \citep{belyaev2015properties,Sutherland2016,Bracamontes-Ramirez2024} we solve the inertial wave equation within a stack of thin layers, coupled by interface conditions, to find the total transmission and reflection coefficients.
The fundamental theory and our models are described in Section \ref{sec:Theory}.
We validate our theoretical calculations via Direct Numerical Simulation using the Dedalus code \citep{Dedalus}. The simulation setup is described in Section \ref{sec:Numerics}, and we compare the theory and simulations in Section \ref{sec:Results}.

\section{Theory and modelling} \label{sec:Theory}
\subsection{General solution for inertial waves in constant geostrophic shear}
\begin{figure}
    \centerline{\includegraphics[width = 0.4\textwidth]{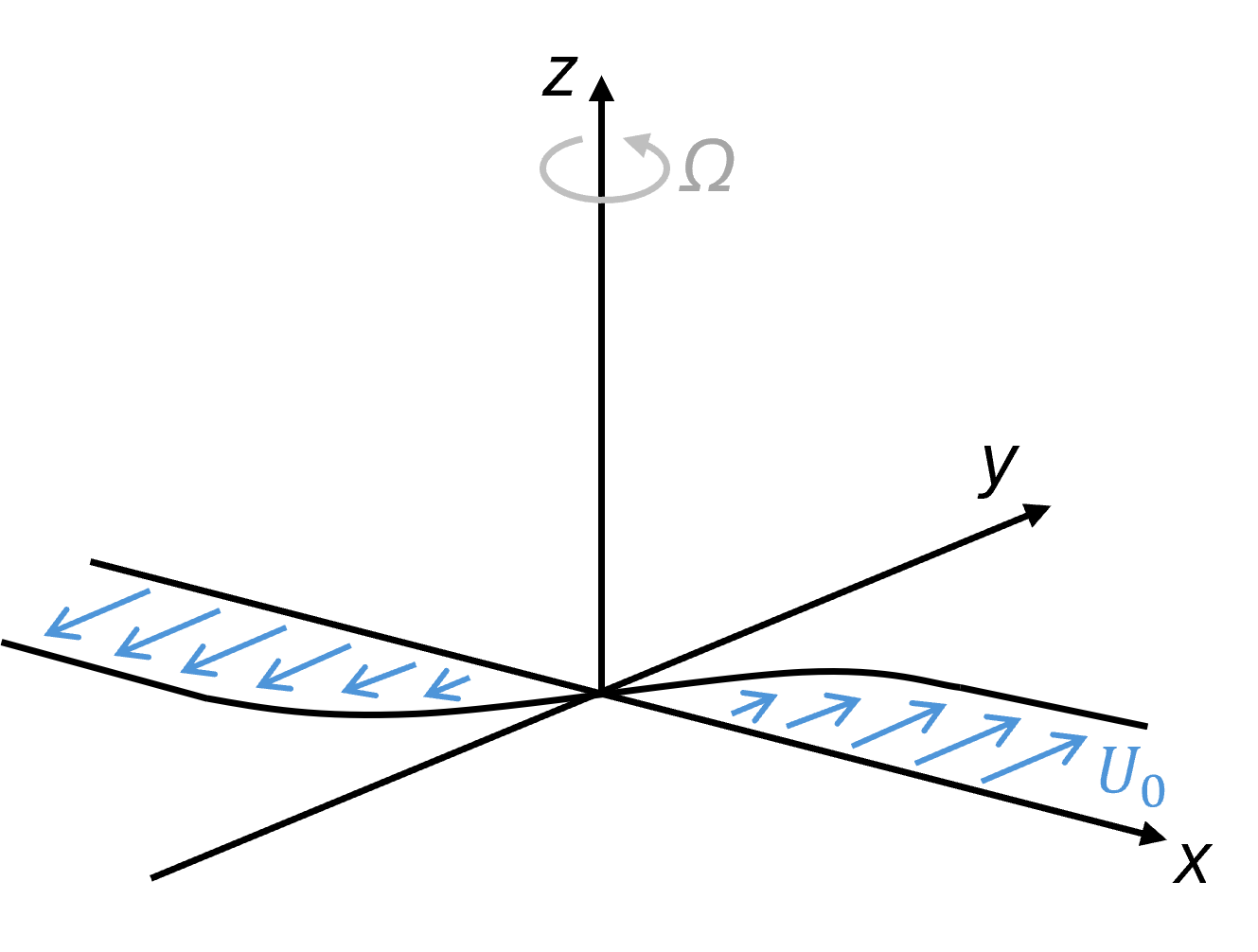}}
    \caption{Schematic figure demonstrating the geometry of the problem.\\ Blue arrows denote an exapmle flow $U_0$ in $y$-direction.}
    \label{fig:Geometry}
\end{figure}

In a rotating frame of reference, with constant angular velocity $\bm \Omega = \Omega\, \bm e_z$ the Navier-Stokes equations for an inviscid and incompressible fluid are
\begin{equation}\label{eq:NS_mom}
    \partial_t \bm u + 2\Omega\,\bm e_z \times \bm u + (\bm u \cdot \nabla)\,\bm u = -\frac{1}{\rho_0}\nabla p
\end{equation}
\begin{equation}\label{eq:NS_con}
    \nabla \cdot \bm u = 0\, ,
\end{equation}
where $\bm u$ is the velocity vector, $\rho_0$ is a constant density and $\nabla p$ is the reduced pressure gradient absorbing the centrifugal force. We are interested in the case where we split up the velocity field into a prescribed geostrophic background flow $\bm U_0 = U_0(x)\,\bm e_y$ and a variation $\bm u(x, z, t)$ from this background state: $\bm u \rightarrow \bm u + \bm U_0$. We assume that the background flow, associated with a reduced background pressure $p_0$, itself satisfies equations \eqref{eq:NS_mom}  and \eqref{eq:NS_con}. Furthermore we make the assumption that all quantities are independent of $y$, such that we treat a quasi-two-dimensional problem.
A schematic of the setup is shown in figure \ref{fig:Geometry}.\\

\noindent We do not make any assumption on the spatial scales of the background flow $U_0$ or the wave field.
We non-dimensionalize the equations with the wavelength of the perturbation $\lambda$ and the rotation rate $\Omega$ using the dimensionless coordinates
\begin{equation}
    (x', z', t') = \left( \frac{x}{\lambda}\,,\, \frac{z}{\lambda}\,,\, \frac{t}{\Omega} \right).
\end{equation}

Then the velocities are non-dimensionalized by $\lambda\Omega$,
\begin{equation}
    \bm u' = \frac{1}{\lambda \Omega}\bm u \hspace{0.25cm}  \text{and} \hspace{0.25cm} U_0' = \frac{1}{\lambda \Omega} U_0. 
\end{equation} 
We assume the perturbation amplitude $|\bm u|$ is small, implying a small Rossby number $Ro=|\bm u|/(\lambda \Omega)\ll1$, such that the non-linear term $\bm u \cdot \nabla \bm u$ is negligible.
We define the dimensionless pressure as
\begin{equation}
    p' = \frac{1}{\rho_0\,(\lambda \Omega)^2}\,p.
\end{equation}
Then the dimensionless equations are
\begin{equation}\label{eq:mom}
    \partial_t' \bm u' + 2\bm e_z \times \bm u' + u'_x\, \partial_x' U_0'\, \bm e_y  = -\nabla p'
\end{equation}
\begin{equation}\label{eq:con}
    \nabla' \cdot \bm u' = 0\, .
\end{equation}
In all further considerations, the prime denoting a dimensionless quantity will be omitted and all variables are dimensionless unless stated otherwise.\\
From equation \eqref{eq:mom}, we can see that the advection term represents a straining of the perturbation $\bm u$ in the $y$-direction. Note that, since we assume $\bm u$ to be independent of $y$, we do not consider advection of the perturbation by the background flow.\\
Eliminating the pressure, taking $\partial_t \nabla \times (\nabla \times \text{\eqref{eq:mom}})$ \citep{davidson2013turbulence}, the Navier-Stokes equation takes the form
\begin{equation}\label{eq:WaveEq}
    \partial^2_t \nabla^2 \bm u + (2\,\bm e_z \cdot \nabla)^2\bm u + \bm S(u_x\,\partial_x U_0) = 0.
\end{equation}
This wave equation includes the new straining term represented by the operator
\begin{equation}
    \bm S = \begin{pmatrix}
                2(\bm e_z \cdot \nabla)\partial_z \\
                (\partial_x^2 + \partial_z^2)\partial_t \\
                -2(\bm e_z \cdot \nabla)\partial_x)
                \end{pmatrix}
\end{equation}
acting on $u_x\, \partial_x U_0$.
This wave equation is valid for arbitrary geostrophic flows $U_0(x)$.\\
Now consider the case of constant geostrophic shear, $\partial_x U_0 = \text{const}$.
Substituting a plane wave ansatz into the first component of equation \eqref{eq:WaveEq}
\begin{equation}\label{eq:Ansatz}
    u_x = \Tilde{u}\,\exp{i(k_x\,x + k_z\,z - \omega t)}
\end{equation}
leads to a modified inertial waves dispersion relation
\begin{equation} \label{eq:kRelation}
    k_x = \pm \gamma k_z, \hspace{1cm} \gamma = \sqrt{\frac{4}{\omega^2} + \frac{2\,\partial_x U_0}{\omega^2}-1}.
\end{equation}
For $\partial_x U_0 = 0$, we recover the classical inertial wave dispersion relation, 
with propagating waves for $\omega < 2$. A non-zero background shear $\partial_x U_0$ controls the maximum frequency 
\begin{equation} \label{omega_max}
   \omega_\text{max} = 2 \sqrt{1 + \frac{\partial_x U_0}{2}}
\end{equation}
for which $\gamma$ is real and classical wave propagation is still possible, as already demonstrated by \citet{kunze1985near}. If the wave frequency $\omega$ exceeds this value, $\gamma$ and $k_x$ become imaginary and the phases only propagate vertically while the entire wave decays exponentially in the horizontal direction. These are evanescent inertial waves as studied by \citet{nosan2021evanescent}.\\
A physical interpretation of the effect of the background shear on the maximum wave frequency is the following. A positive value of $\partial_x U_0$ represents a positive background vorticity, which enhances the restoring force of the Coriolis acceleration. A stronger restoring force supports higher frequencies in an oscillatory system and thus, $\omega_\text{max} > 2$. In contrast, a negative background vorticity weakens the restoring force, which fails to support high frequency motion, where $\omega_\text{max} < 2$.\\
For a given vertical wave number $k_z$ and wave frequency $\omega$,
we have the polarization relations
\begin{equation}
    u_y = -i\frac{2 + \partial_x U_0}{\omega}\,u_x \hspace{0.5cm} \text{and} \hspace{0.5cm} u_z = -\frac{k_x}{k_z}\,u_x.
\end{equation}
Finally we can obtain the pressure $p$ from integrating the third component of the momentum equation \eqref{eq:mom} as
\begin{equation}
    p(x,z) = - \frac{\omega\, k_x}{k_z^2}\,u_x.
\end{equation}
Thus, we obtain a full analytical description of inertial waves and evanescent inertial waves inside any constant geostrophic shear $\partial_x U_0$.

\subsection{Interaction of inertial waves with a localized geostrophic shear layer}
\begin{figure}
  \centerline{\includegraphics[width = 0.3\textwidth]{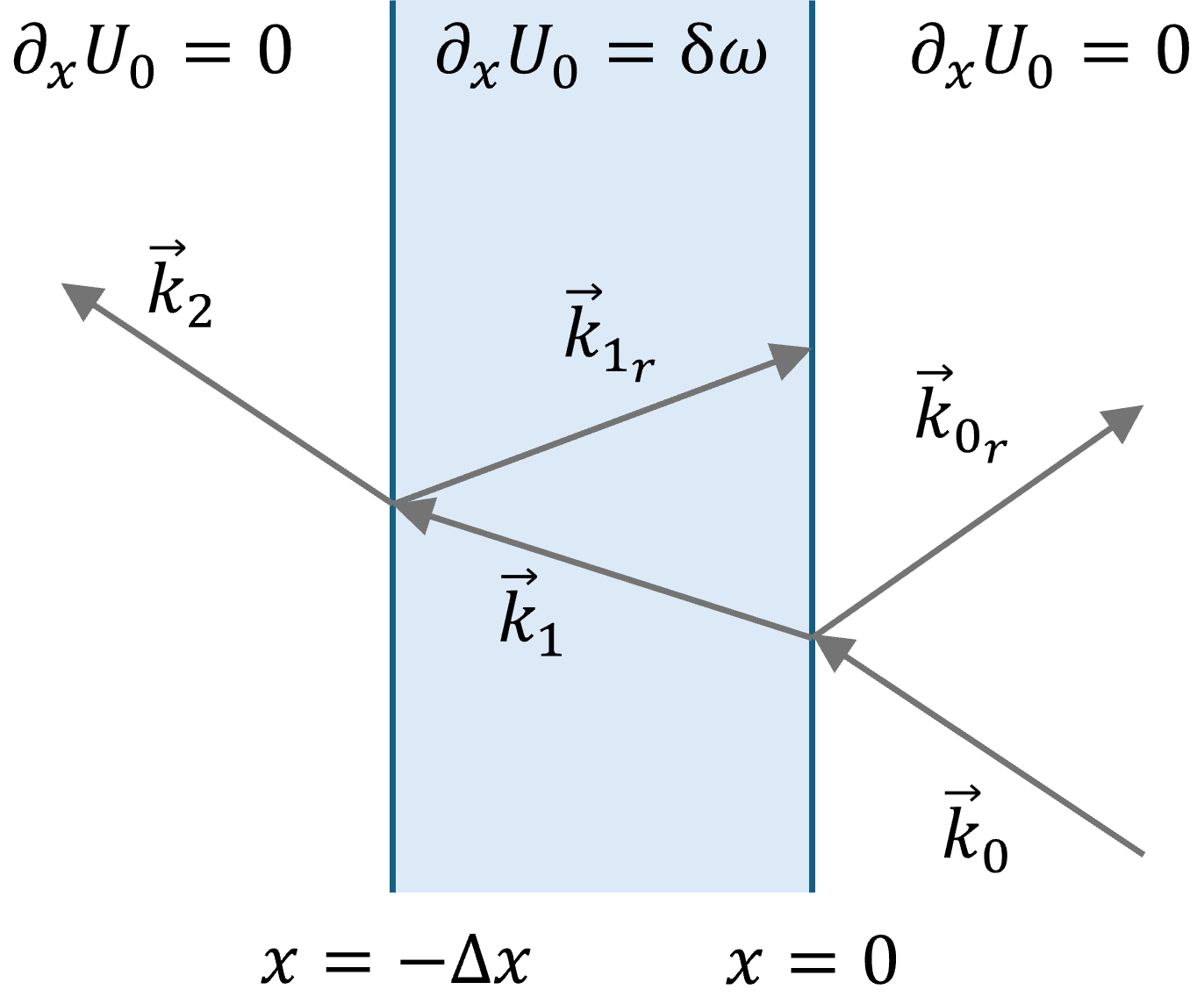}}
  \caption{Model of plane inertial waves interacting with a single layer of constant shear.\\ The wave vectors $\bm k_{i_{(r)}} = (k_{i_{(r)}, x},\, k_{i_{(r)}, z})$ are indicated schematically.}
\label{fig:SingleLayer_Model}
\end{figure}

We now consider inertial waves propagating into a localized layer of constant geostrophic shear, as depicted in figure \ref{fig:SingleLayer_Model}.

\noindent We assume the incident wave has amplitude, wavenumber, and frequency $(\Tilde{u}_0, \bm k_0, \omega_0)$, which we assume to be known. Our aim is to describe the reflected waves $(\Tilde{u}_{0_r}, \bm k_{0_r}, \omega_{0_r})$ \& $(\Tilde{u}_{1_r}, \bm k_{1_r}, \omega_{1_r})$, and the transmitted waves $(\Tilde{u}_1, \bm k_1, \omega_1)$ \& $(\Tilde{u}_2, \bm k_2, \omega_2)$. 
For this we follow a common procedure for inertial/internal waves at diverse interfaces (e.g., \citealt{phillips1963energy, sutherland2004internal, belyaev2015properties}): we impose continuity conditions for the pressure and the flow component normal to the interfaces at $x = 0$ and $x = -\Delta x$.
\begin{alignat}{3}
    u_{1,x} + u_{1_r, x} \quad & = \quad u_{2,x} \quad && \vert_{x = -\Delta x} \label{eq:cont_u1}\\
    p_1 + p_{1_r}        \quad & = \quad p_2          \quad && \vert_{x = -\Delta x} \\
    u_{0,x} + u_{0_r, x} \quad & = \quad u_{1,x} + u_{1_r,x} \quad && \vert_{x = 0} \\
    p_0 + p_{0_r}        \quad & = \quad p_1 + p_{1_r} \quad && \vert_{x = 0} \label{eq:cont_p0}
\end{alignat}
As the problem is autonomous in $z$ and $t$, all waves must possess the same vertical wavenumber $k_z$ and the same frequency $\omega$.
This means the only free variables left are the horizontal wave numbers $k_{i_{(r)},x}$ and the amplitudes $\Tilde{u}_{i_{(r)}}$  ($i = 1,2$ and the index '$(r)$' stands for both transmitted and reflected wave).\\
For a given strength of the geostrophic shear $\partial_xU_0=\delta\omega$ and imposed ($\Tilde{u}_0, k_{x,0}, k_{z},\,\omega$) by the incident wave, we can calculate the horizontal wavenumbers in every part of the domain using equation \eqref{eq:kRelation}. This leaves us---by inserting the ansatz \eqref{eq:Ansatz} into the continuity conditions \eqref{eq:cont_u1}-\eqref{eq:cont_p0}---with four linear equations for four unknown amplitudes
\begin{alignat}{3}
    \Tilde{u}_{1}\, \exp{(-ik_{1,x}\,\Delta x)} + \Tilde{u}_{1_r}\, \exp{(-ik_{1_r,x}\,\Delta x)} \, & = \,\Tilde{u}_{2}\, \exp{(-ik_{2,x}\,\Delta x)} \\
    k_{1,x}\Tilde{u}_{1}\, \exp{(-ik_{1,x}\,\Delta x)} + k_{1_r,x}\Tilde{u}_{1_r}\, \exp{(-ik_{1_r,x}\,\Delta x)}       \, & =  \, k_{2,x}\Tilde{u}_{2}\, \exp{(-ik_{2,x}\,\Delta x)}  \\
    \Tilde{u}_{0} + \Tilde{u}_{0_r} \, & = \, \Tilde{u}_{1} + \Tilde{u}_{1r}\\
    k_{0,x}\Tilde{u}_{0} + k_{0_r,x}\Tilde{u}_{0_r} \, & = \, k_{1,x}\Tilde{u}_{1} + k_{1_r,x}\Tilde{u}_{1r}
\end{alignat}
Solving for the unknown amplitudes we obtain:
\begin{alignat}{3}
    &\Tilde{u}_{0_r} \quad  &=& \quad \alpha_0 \,\Tilde{u}_0, \\
    &\Tilde{u}_{1} \quad  &=& \quad \beta_0 \, \Tilde{u}_0, \\
    &\Tilde{u}_{1_r} \quad   &=& \quad \alpha_1\,\Tilde{u}_1, \\
    &\Tilde{u}_2 \quad  &=& \quad \beta_1\,\Tilde{u}_1, 
\end{alignat}
with
\begin{alignat}{3}
    \alpha_1 & = \quad \frac{k_{2,x} - k_{1,x}}{k_{1_r,x} - k_{2,x}}\exp[-i(k_{1,x} - k_{1_r,x})\Delta x] \label{eq:R1}\\
    \beta_1 & = \quad \left(\exp[-ik_{1,x}\Delta x] + \alpha_1\exp[-ik_{1_r,x}\Delta x]\right)\exp[ik_{2,x}\Delta x] \label{eq:T1}\\
    \alpha_0 & = \quad \frac{k_{1,x} - k_{0,x} + \alpha_1(k_{1_r,x} - k_{0,x})}{k_{0_r,x} - k_{1,x} + \alpha_1(k_{0_r,x} - k_{1_r,x})}\label{eq:R0}\\
    \beta_0 & = \quad \frac{1 + \alpha_0}{1 + \alpha_1}\label{eq:T0}
\end{alignat}

\noindent Note that for the case of an unperturbed background vorticity, i.e. $\delta \omega = 0$, we must recover the uniform propagation of inertial waves, i.e. $\alpha_1 = \alpha_0 = 0$ and $\beta_1 = \beta_0 = 1$. This allows to determine the signs of the horizontal wave vectors. As their magnitudes are equal, to satisfy $\alpha_0=0$ we have
\begin{equation}
    \text{sign}(k_{1,x}) = \text{sign}(k_{0,x}) \hspace{0.5cm} \text{and} \hspace{0.5cm} \text{sign}(k_{0_r,x}) = -\text{sign}(k_{0,x})
\end{equation}
and to satisfy $\alpha_1=0$ we have
\begin{equation}
    \text{sign}(k_{2,x}) = \text{sign}(k_{0,x}) \hspace{0.5cm} \text{and} \hspace{0.5cm} \text{sign}(k_{1_r,x}) = -\text{sign}(k_{0,x})
\end{equation}
In other words, we obtain the expected result that the transmitted phases travel in the same horizontal direction as the incoming waves, while the reflected phases travel in the opposite direction.\\
At this point, we introduce the concept of critical shear. We define it as the value $\partial_x U_0 = \delta \omega_c$ below which $\gamma$ becomes imaginary inside the shear layer for an inertial wave of frequency $\omega$. 
We can obtain this value by solving equation \eqref{omega_max} for $\partial_x U_0$ and setting $\omega_\text{max} = \omega$.
\begin{equation} \label{eq:critShear}
    \delta \omega_c = \frac{\omega^2}{2} - 2
\end{equation}
For our problem, this critical shear is a negative number (as $\omega \leq 2$) and we define shear layers where $\delta \omega < \delta \omega_c$ as supercritical shear layers.\\
Hence, in supercritical shear layers $k_x$ is imaginary. As we expect evanescent inertial waves to always decay towards the inside of impenetrable media, we impose
\begin{equation}
    k_{1,x} = -i|\gamma k_z| \hspace{0.5cm} \text{and} \hspace{0.5cm} k_{1_r,x} = - k_{1,x}, \hspace{0.5cm} \text{if} \hspace{0.3cm} \delta \omega < \delta \omega_c < 0.
\end{equation}
This ensures that the wave $\bm u_1$, transmitted into the layer of supercritical shear, decays in negative $x$-direction, i.e., towards the inside of the layer.

\subsection{Interaction with arbitrary piecewise-constant geostrophic shear profiles}
\begin{figure}
  \centerline{\includegraphics[width=0.8\textwidth]{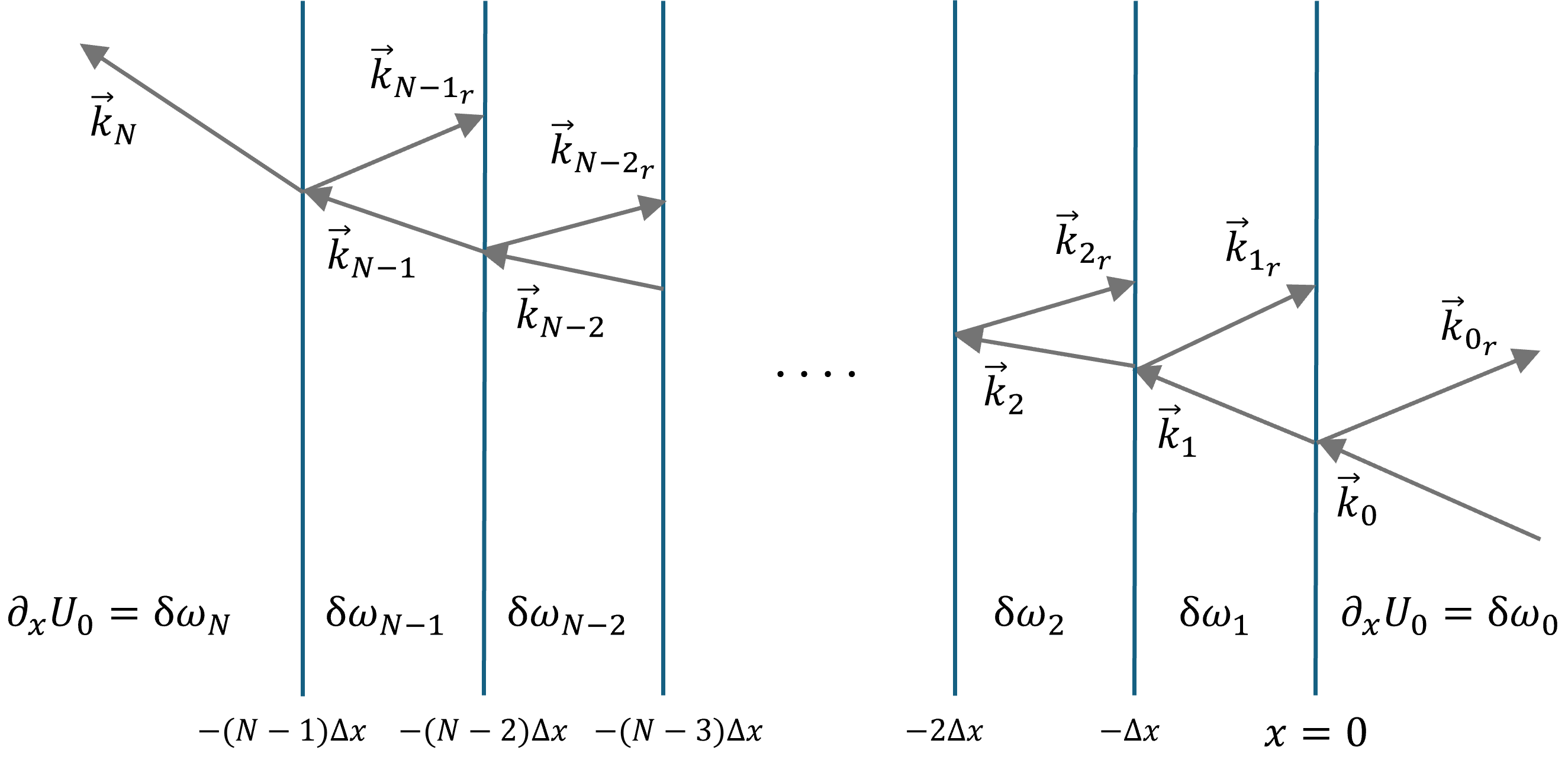}}
  \caption{Model of multiple consecutive shear layers traversed by plane inertial waves. \\ This expands the model of a single layer and two interfaces to a case of $N$ interfaces.}
\label{fig:MultiLayer_Model}
\end{figure}

Let us now consider the scenario where an inertial wave 
propagates into a series of shear layers, each with width $\Delta x$ and shear strength $\delta \omega_i$,
i.e., a piecewise constant shear profile, as illustrated in figure \ref{fig:MultiLayer_Model}. As we shall see, by adjusting the width and amplitudes of the shear layers we can use such a model to approximate arbitrary geostrophic shear profiles.

\noindent Imposing the continuity conditions for $p$ and $u_x$ at each interface, we obtain the following recursive relations at the $n$-th interface ($n = 0,..., N-1$)
\begin{align}
 \begin{split}
    \Tilde{u}_{n+1}\,a_{n+1} + \Tilde{u}_{n+1_r}\,a_{n+1_r} &= \Tilde{u}_{n}\,b_{n} + \Tilde{u}_{n_r}\,b_{n_r} \\
    k_{n+1, x}\Tilde{u}_{n+1}\,a_{n+1} + k_{n+1_r, x}\Tilde{u}_{n+1_r}\,a_{n+1_r} &= k_{n,x}\Tilde{u}_{n}\,b_{n} + k_{n_r,x}\Tilde{u}_{n_r}\,b_{n_r}, 
 \end{split}
 \end{align}
 where 
\begin{align}
 \begin{split}
     a_{n_{(r)}} &= \exp{[(-ik_{n_{(r)},x})\,(n-1)\Delta x]} \\
     b_{n_{(r)}} &= \exp{[(-ik_{n_{(r)},x})\,n\Delta x]} \\
 \end{split}
\end{align}
are phase factors resulting from the offset of the $n$-th interphase from $x = 0$.\\
Taking the incident amplitude as $\Tilde{u}_0 = Ro$ and setting $\Tilde{u}_{N_r} = 0$, we can formulate the following system of equations. \vspace{0.5cm}
\[
\begin{pmatrix}
1 & 0 & 0 & \cdots & 0 & 0 \\
-b_0 & -b_{0_r} & a_1 & a_{1_r} & 0 & \vdots \\
-k_0\,b_0 & -k_{0_r}\,b_{0_r} & k_1\,a_{1} & k_{1_r}\,a_{1_r} & 0 & \vdots \\
0 & -b_1 & -b_{1_r} & a_2 & a_{2_r} & \vdots \\
0 & -k_1\,b_1 & -k_{1_r}\,b_{1_r} & k_2\,a_{2} & k_{2_r}\,a_{2_r} & \vdots \\
\vdots & 0 & \ddots & \ddots & & \\
\vdots & & -b_{N-1} & -b_{N-1_r} & a_N & a_{N_r} \\
\vdots & & -k_{N-1}\,b_{N-1} & -k_{N-1_r}\,b_{N-1_r} & k_N\,a_{N} & k_{N_r}\,a_{N_r} \\
0 & \cdots & 0 & 0 & 0 & 1
\end{pmatrix}
\begin{pmatrix}
\tilde{u}_0 \\[6pt]
\tilde{u}_{0_r} \\[6pt]
\tilde{u}_1 \\[6pt]
\tilde{u}_{1_r} \\[6pt]
\vdots \\[4pt]
\tilde{u}_{N-1} \\[6pt]
\tilde{u}_{N-1_r} \\[6pt]
\tilde{u}_N \\[2pt]
\tilde{u}_{N_r}
\end{pmatrix}
=
\begin{pmatrix}
Ro \\[6pt]
0 \\[6pt]
0 \\[6pt]
0 \\[6pt]
\vdots \\[4pt]
0 \\[6pt]
0 \\[6pt]
0 \\[2pt]
0
\end{pmatrix}
\]

\vspace{0.5cm}

\noindent We solve this system and obtain the reflected amplitude $\tilde{u}_{0_r}$ and the transmitted amplitude $\tilde{u}_N$.\\

\begin{figure}
  \centerline{\includegraphics[width=0.85\textwidth]{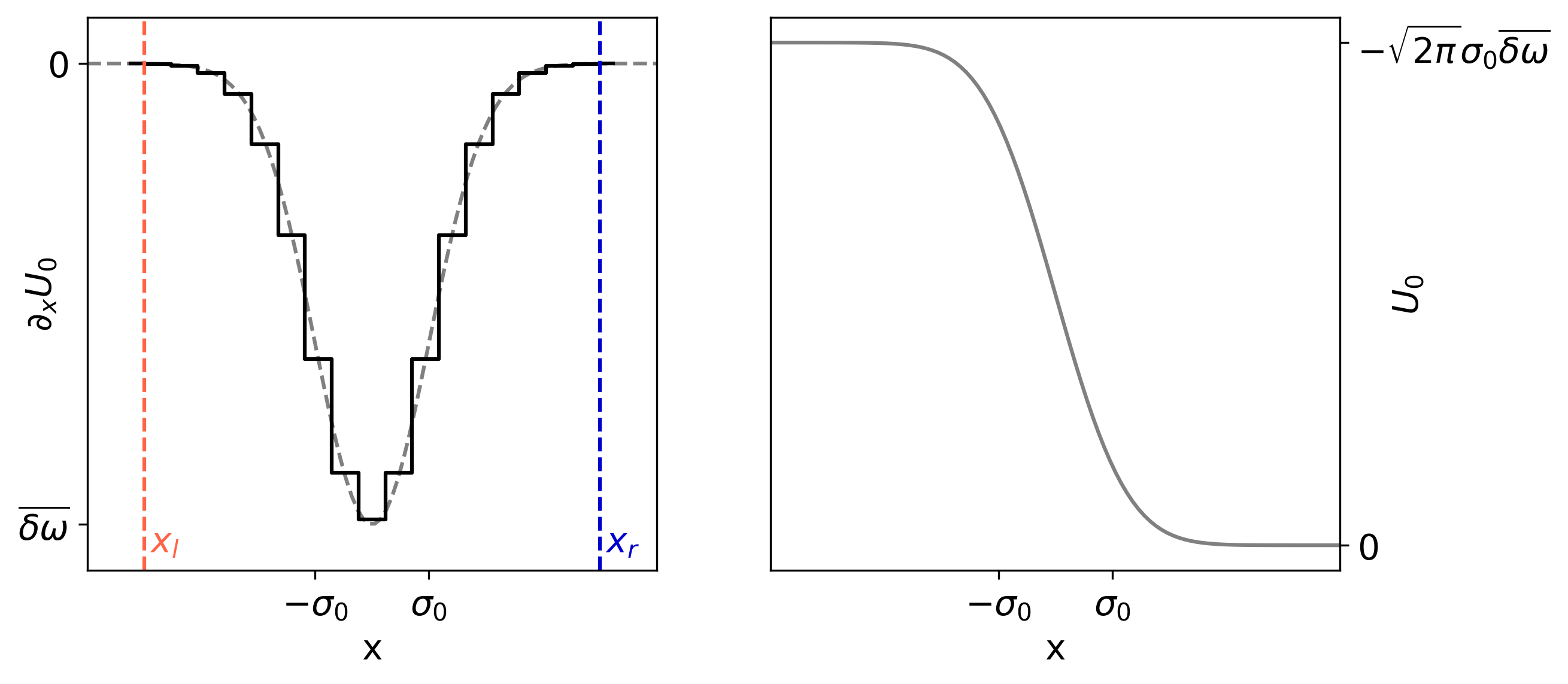}}
  \caption{Left: Segmentation of a Gaussian shear profile using $N = 18$ interfaces. Right: The flow profile corresponding to the shear profile, i.e., a shear zone in the shape of an \textit{erfc} centred around $x = 0$.}
\label{fig:Segmentation}
\end{figure}

Any continuous geostrophic profile can be approximated by this type of piecewise constant shear profile. 
As an example for our problem, consider the following Gaussian shear profile:
\begin{equation} \label{eq:GaussShear}
    \partial_x U_0(x) = \overline{\delta \omega}\, \exp{(-\frac{x^2}{2\sigma_0^2})}
\end{equation}
Our strategy of discretizing this profile is to define an interval $[x_l, x_r] = [-4\,\sigma_0,  4\,\sigma_0]$ outside of which we assume $\partial_x U_0(x) = 0$. We define $N$ equidistant interfaces within this interval, creating $N-1$ layers of equal thickness $\Delta x$. We define the $\delta \omega_i$ ($i = 1,\dots , N-1$) as the mean value of $\partial_x U_0$ within each layer. 
In figure \ref{fig:Segmentation} we show an example discretization with $N=18$.
However, to obtain a better approximation of the profile we can increase $N$ arbitrarily. For convenience, we will henceforth denote the transmitted wave with an index '$t$' when comparing the piecewise-constant model to the continuous case: e.g., $\tilde{u}_N \rightarrow \tilde{u}_t$.

\noindent Finally, the calculation described above assumes the first shear layer is at $x=0$. However, the calculation of the reflection and transmission coefficients is independent of the position of the first shear layer, so the same approach can be used if the first shear layer is at another position (e.g., $x_r$ as in figure \ref{fig:Segmentation}). Such a translation simply requires a change of the phase factors $a_{n_{(r)}}$ and $b_{n_{(r)}}$.

\section{Numerical Methods} \label{sec:Numerics}

\begin{figure}
  \centerline{\includegraphics[width=\linewidth]{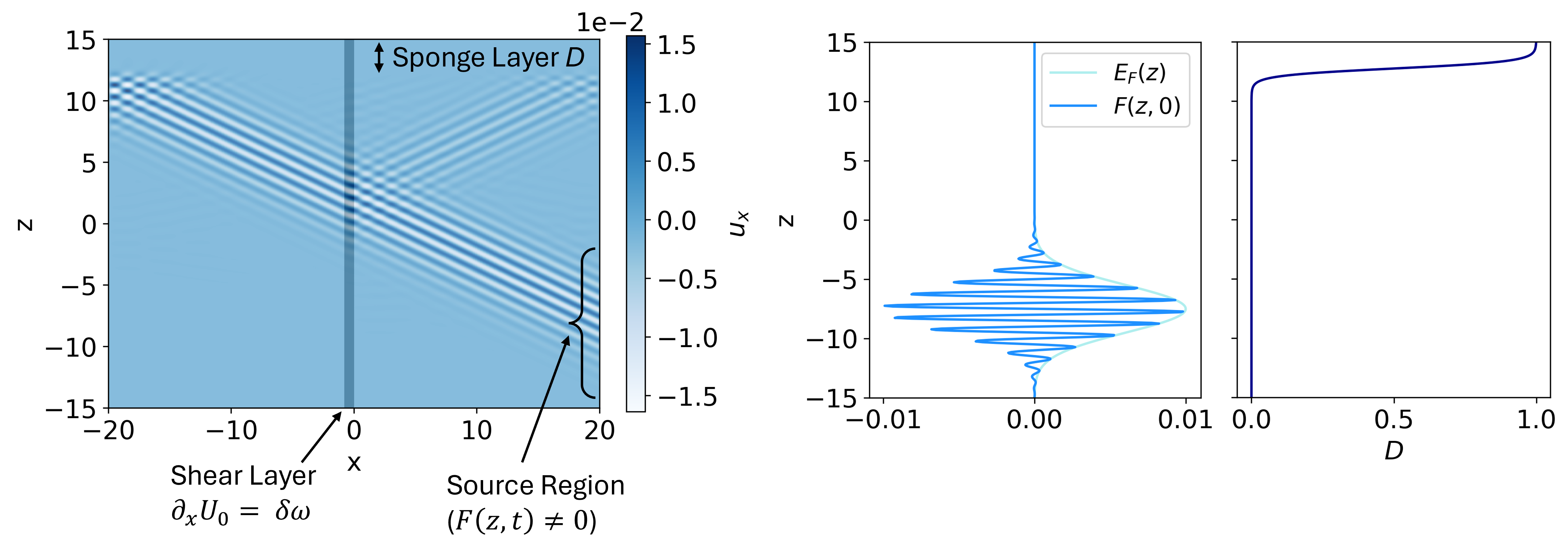}}
  \caption{Typical setup for the DNS using Dedalus. A shear layer is in the middle of the domain. Energy is injected on the right boundary using the source function $F$---a sine wave modulated by a Gaussian envelope $E_F$. A sponge layer $D$ at the top acts as an energy sink.}
\label{fig:Dedalus_Setup}
\end{figure}

To validate our theory we run Direct Numerical Simulations (DNS) using Dedalus \citep{Dedalus}. Dedalus is a spectral solver for partial differential equations. We use it to solve the equations \eqref{eq:NS_mom} and \eqref{eq:NS_con} by expanding all variables as a Chebyshev series in $x$ and Fourier series in $z$. In the spectral expansion we utilize both $1024$ Chebyshev polynomials and $1024$ Fourier modes for a domain of size $40 \times 30$.\\ 
We specify a set of boundary and initial conditions in a typical setup suited for the reflection-transmission problem for inertial waves, which is shown in figure \ref{fig:Dedalus_Setup}.

\noindent We excite the waves on the right boundary by imposing
\begin{equation}\label{eq:Source}
    u_x(x = 20, z, t) = F(z,t) = Ro\, \exp{\left(-\frac{(z-z_\omega)^2}{2d_\omega^2}\right)}\, \sin(k_z z - \omega t),
\end{equation}
where $z_\omega = -7.5$ and $Ro = 0.01$.
We choose the vertical wavelength $\lambda_z=1$ such that $k_z=2\pi$.
The extent of the source region is chosen as $d_\omega = 2$. We furthermore always impose the frequency $\omega = 1.8$.\\
We use a no-penetration boundary condition on the left boundary, i.e.,
\begin{equation}
    u_x(x = -20, z, t) = 0,
\end{equation}
such that both right and left boundaries are reflecting.
The top and bottom boundaries are periodic. As an initial condition we choose 
\begin{equation}
    \bm u(\bm x, t = 0) = p(\bm x, t = 0) = 0.
\end{equation}
To avoid a steady increase of energy in the system, we place a sponge layer at the top of the domain by adding an extra damping term $-\bm u D(z)$ on the right hand side of the momentum equation \eqref{eq:WaveEq}, with
\begin{equation}
    D(z) = \frac{1}{2}\left[\tanh{\left(\frac{z - z_D}{d_D}\right)} + 1\right].
\end{equation}
$z_D = 12.75$ is the vertical position of the sponge layer with width $d_D = 0.75$. Although $D$ is not a periodic function, we do not expect any significant effect on our results caused by its slight misrepresentation in the Fourier basis. Only waves travelling downward or through the layer would be affected. To be sure, we have performed a single simulation for a larger domain and a periodic function, i.e. by adding a second absorbing layer at $z = -45$ (the increase in domain size is to prevent an interference of the lower layer with the source domain), obtaining equivalent results to those when using the non-periodic sponge layer.\\
We run our simulations until $t = 800$ in order to reach a steady wave pattern. 
Typically, one simulation takes 20 minutes, running on four cores of a 'Core i9-14900KF' produced by \textit{Intel}.\\
While our analytic theory is for plane waves, the simulations are of wave beams (see, e.g., \citealt{le2020fluid}). However, in the centre of the beams we assume the inertial waves behave as plane waves and will thus consider the central peak of the beams for a comparison to our theory.\\

\begin{figure}[ht!]
    \centerline{\includegraphics[width=\linewidth]{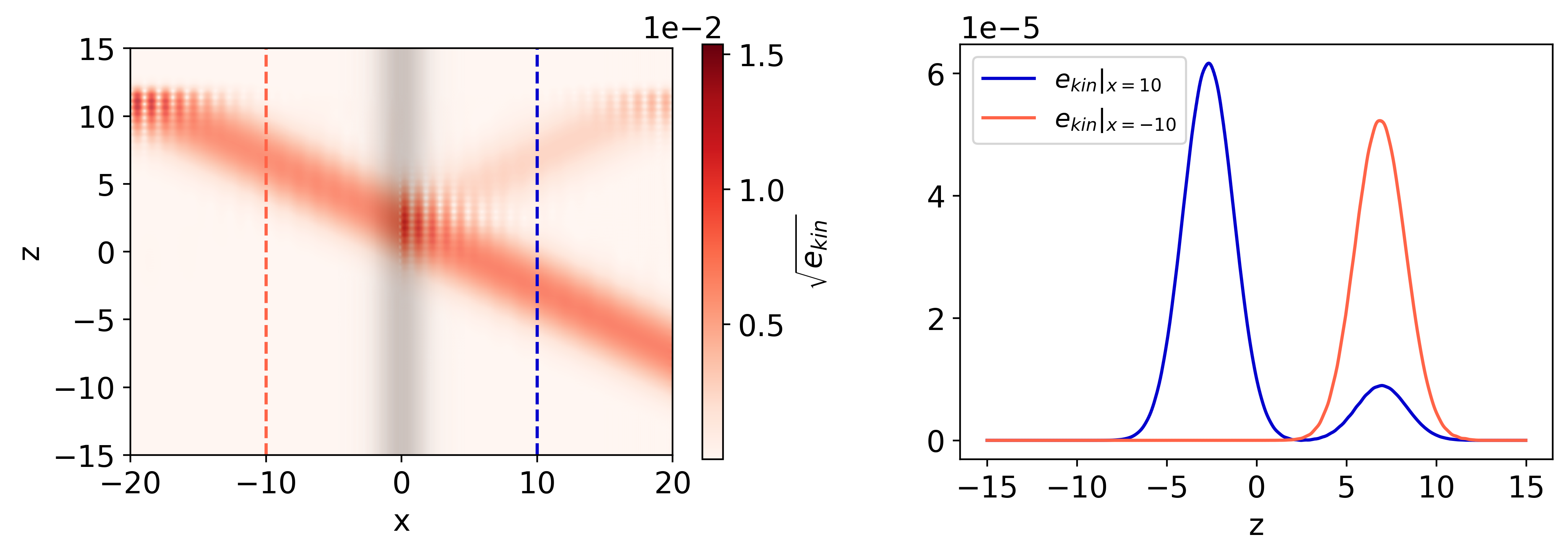}}
    \caption{Left: Kinetic energy field---taking the square root allows for less discrepancy between strong and weak signals. The red and blue line show the profiles along which we measure the kinetic energy. Right: Vertical profiles of the kinetic energy.}
    \label{fig:KE_Method}
\end{figure}

The kinetic energy density of a wave field is defined as 
\begin{equation}
    e_{\text{kin}} = \frac{1}{2} Re(\bm u) \boldsymbol{\cdot} Re(\bm u).
\end{equation}
In regions of zero shear (for $\partial_x U_0 = 0$) the kinetic energies of the plane inertial waves is
\begin{equation}
    e_{\text{kin}} = |\Tilde{u}|^2\frac{2\Omega^2}{\omega^2}.\label{eq:ekin}
\end{equation}
As we impose $u_{0,x}$ on the right boundary with a peak amplitude of $Ro = 0.01$, the peak kinetic energy of the incident wave will always be
\begin{equation}
    e_{\text{kin},0} = Ro^2\frac{2\Omega^2}{\omega^2} \approx 6.17\times 10^{-5}.
\end{equation}
We compare the transmitted and reflected waves to our analytic predictions in terms of the transmitted and reflected kinetic energy.\\
To estimate the kinetic energy in the DNS we proceed as follows. Given a geostrophic shear localized near $x=0$, we plot the kinetic energy \eqref{eq:ekin} along two vertical lines at $x=10$ and $x=-10$ in the incident and transmitted domains, respectively (see figure \ref{fig:KE_Method}). 
The first peak at $x = 10$ provides the kinetic energy $e_{\text{kin},0}$ in the centre of the incident beam and the second one the kinetic energy in the centre of the reflected beam, $e_{\text{kin},0_r}$. The maximum of the profile at $x = -10$ provides the kinetic energy $e_{\text{kin},2}$ of the transmitted wave.

\section{Results and Discussion} \label{sec:Results} 

\subsection{Localized, 'ideal' shear layer}
\begin{figure}[ht!]
    \centerline{
    \begin{minipage}{\textwidth}
        \includegraphics[width=0.495\linewidth]{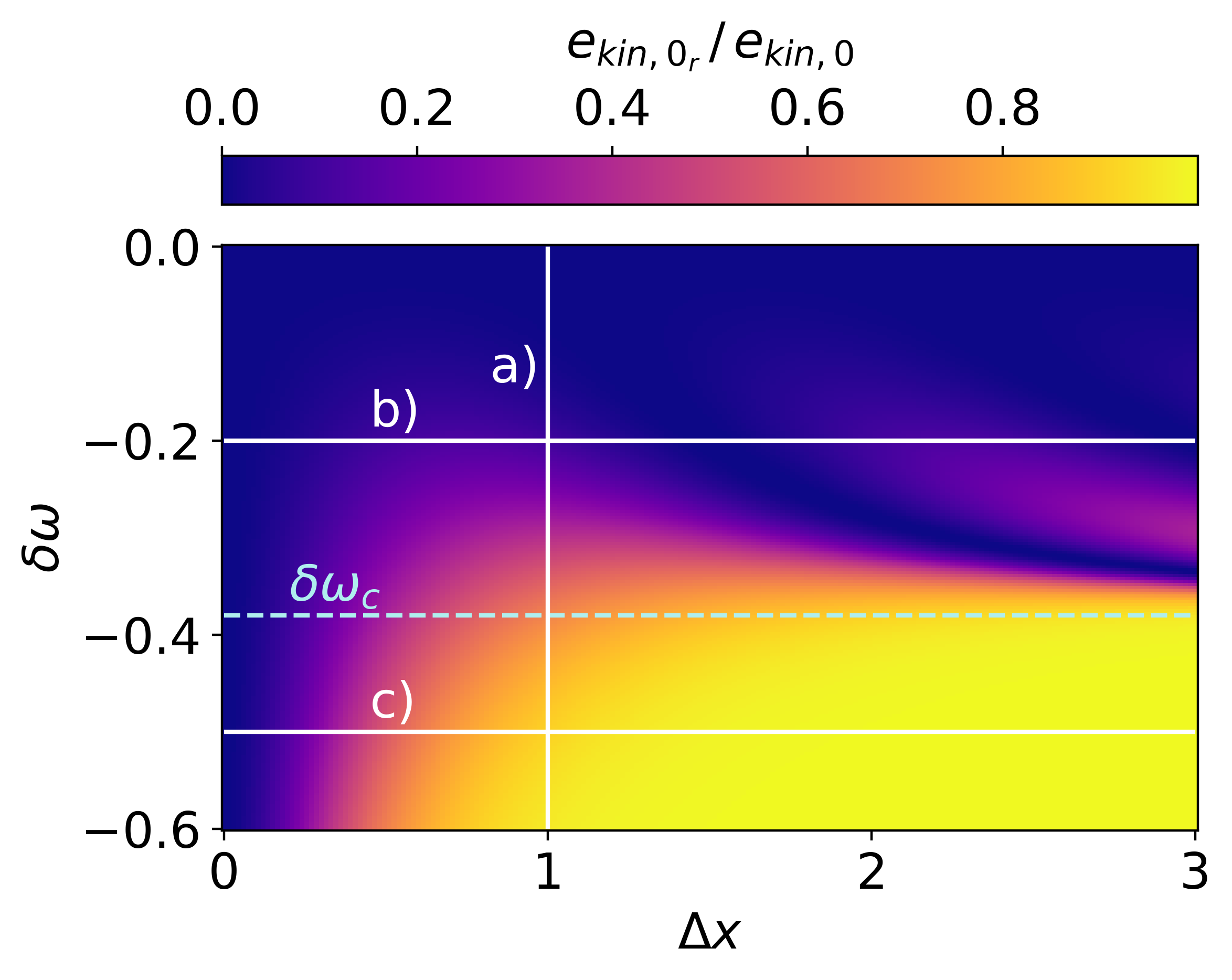}
        \hspace{0.02\textwidth}
        \includegraphics[width=0.47\linewidth]{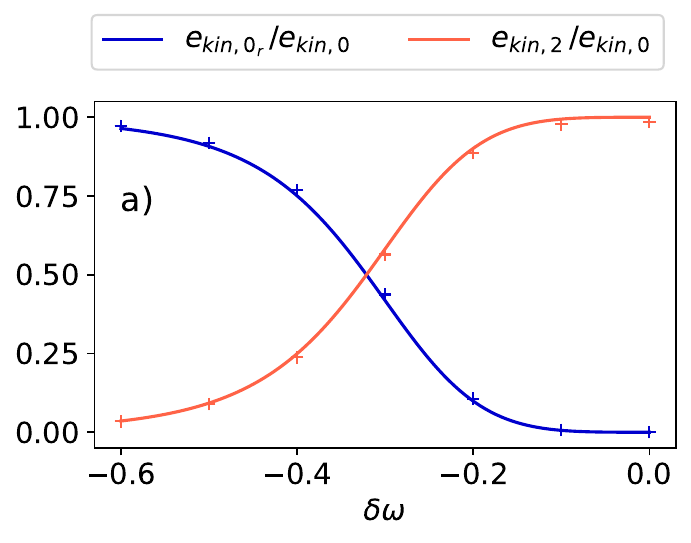}
    \end{minipage}}
    \centerline{
    \begin{minipage}{\textwidth}
        \hspace{0.02\textwidth}
        \includegraphics[width=0.47\linewidth]{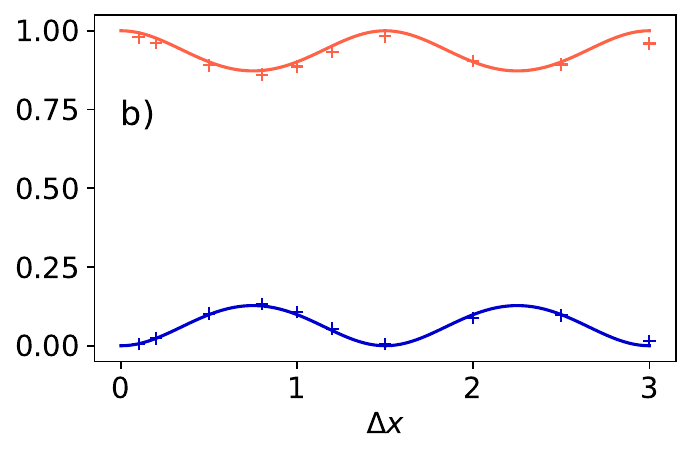}
        \hspace{0.04\textwidth}
        \includegraphics[width=0.47\linewidth]{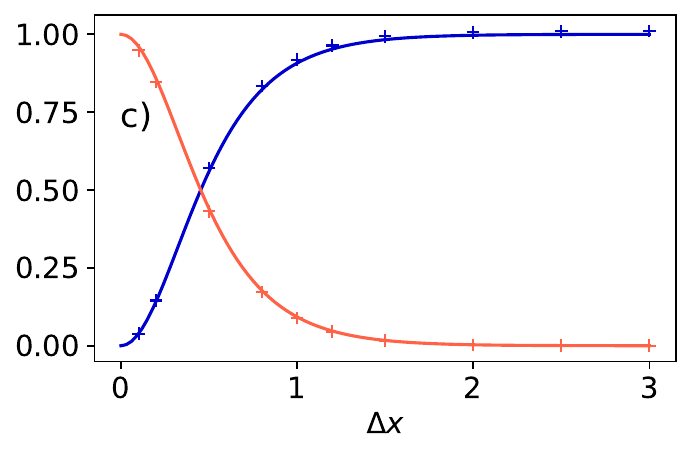}
    \end{minipage}}
    \caption{Predicted and measured reflection and transmission coefficients in the case of a single shear layer. Upper left: Map of the predicted reflection coefficient using the theoretical model for various values of the shear layer thickness $\Delta x$ and shear values $\delta \omega$. The critical shear $\delta \omega_c$ is represented by the dashed line. We choose three profiles, represented by white lines, along which we plot the reflection an transmission coefficients in the remainig panels: a) we keep the layer width constant at $\Delta x = 1$ and vary the shear strength $\delta \omega$, b) and c) we keep $\delta \omega$ constant at -0.2 and -0.5, respectively, and vary $\Delta x$. Solid lines represent theoretical predictions and crosses represent measured values from the DNS.}
\label{fig:SingleLayer_Results}
\end{figure}

We first compare our theory to the simulations in the case of the single discrete shear layer. In figure \ref{fig:SingleLayer_Results} (upper left) we show a map of the energy reflection coefficient $\frac{e_{\text{kin},0_r}}{e_{\text{kin},0}}$ for varying thicknesses and strengths of the shear layer.
We also plot three profiles of reflection and transmission coefficients at either fixed $\delta \omega$ or fixed $\Delta x$ in figure \ref{fig:SingleLayer_Results} (panels a to c). For these profiles, we compare the theory (solid lines) to simulations (symbols), finding good agreement.\\
Two interesting features are: (1) The theory predicts significant back-scattering of energy even for subcritical shear values $\delta \omega > \delta \omega_c$. (2) Even for supercritical shear, the reflection coefficient vanishes for thin enough shear layers.

\noindent Regarding the former observation,
Equation \eqref{eq:R0} shows that any difference between $k_{x,1}$ and $k_{x,0}$ gives a non-zero reflection coefficient. This is because the pressure variation associated to inertial waves is related to the product $k_x \, \Tilde{u}_x$, which decreases for increasingly horizontal waves (smaller $|k_x|$). Hence, if $|k_{x,1}|$ is reduced by a shear flow inside the anomalous layer (Equation \eqref{eq:kRelation}), then the horizontal flow must increase to sustain pressure continuity at the interface. However, an additional horizontal flow only on one side of the interface would violate the continuity of $u_x$ 
requiring the presence of a reflected wave. That is, as soon as a shear flow is introduced, $k_{x,1} \neq k_{x,0}$ and the continuity of $p$ and $u_x$ can only be upheld by both a transmitted and reflected wave.\\
Next note that the reflection coefficient has a periodic dependence on the layer width for fixed subcritical shear strengths. This is because of the interference of the wave reflected at the rear interface, $x = -\Delta x$, and the wave reflected at the frontal interface, $x = 0$. Depending on $k_{x,1}$ (which depends on the shear strength), different widths of shear layers will result in positive or negative interference of the two reflected waves.\\ 
Let us consider the example of profile b) in figure \ref{fig:SingleLayer_Results}, where $\delta \omega = -0.2$ and we have $\gamma = \frac{1}{3}$ such that the horizontal wave length is $\lambda_x = 3$.
In the profile, we can see that the reflected energy has maxima at odd multiples of $\Delta x = \frac{\lambda_x}{4}$.
This is indicating that the two reflected waves, one introduced by a pressure drop at the frontal interface and the other by a pressure increase at the rear interface, interfere positively. The difference in polarity of the pressure changes introduces a phase shift of $\pi$. In addition, we obtain an additional phase shift of $\pi$ when the horizontal path difference, $d = 2\Delta x$, of the waves is an odd multiple of $\frac{\lambda_x}{2}$. Hence, the observed positive interferences (a phase shift of $2\pi$) appear for the expected layer widths of $\Delta x = \frac{d}{2} = n\,\frac{\lambda_x}{4}$, where $n$ is an odd integer.\\
Our calculations also show the transmission of waves even with supercritical shear. This is due to the tunnelling of evanescent waves.
As discussed above, $k_{1,x}$ becomes imaginary in the supercritical case of $\delta \omega < \delta \omega_c$. This means only evanescent waves exist inside the supercritical shear layer. 
The evanescent waves are oscillatory in the vertical and exponentially decaying in the horizontal as they propagate into the supercritical shear layer (see \citealt{nosan2021evanescent}).
However, for thin enough shear layers, the waves do not decay entirely inside the layer and the vertically travelling phases at the rear end of shear the layer excite a propagating wave beyond it. That is, despite the evanescence of the waves inside the layer, energy transfers across the layer through this tunnelling mechanism, similar to what is observed for internal gravity waves at thin mixed layers \citep{sutherland2004internal}. Furthermore, we know that the decay factor is $|\text{Im}(k_{1,x})|$, which increases with the shear strength. Consequently, substantial tunnelling is limited to ever thinner shear layers for decreasing $\delta \omega$, consistent with the results shown in figure \ref{fig:SingleLayer_Results}.

\subsection{Continuous shear profile and layered model} 
\begin{figure}[ht!]
    \centerline{
    \begin{minipage}{\textwidth}
        \includegraphics[width=0.505\linewidth]{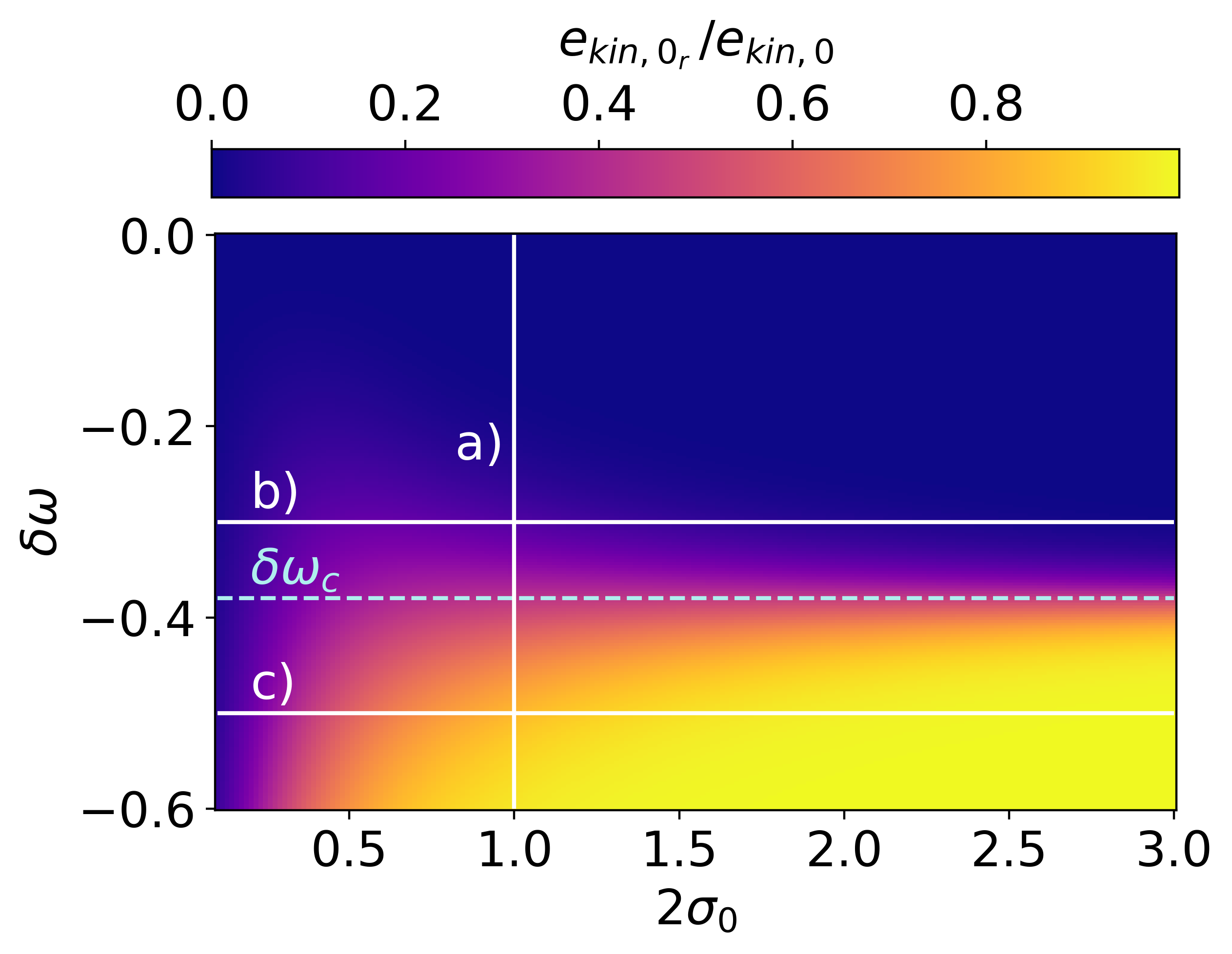}
        \hspace{0.02\textwidth}
        \includegraphics[width=0.46\linewidth]{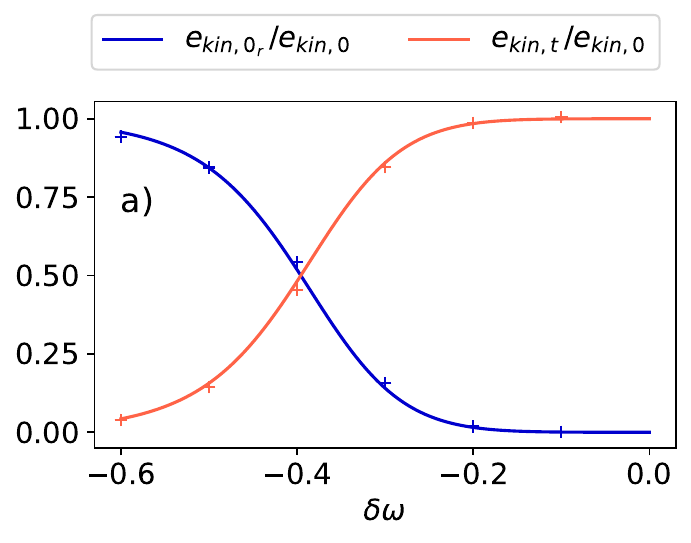}
    \end{minipage}}
    \centerline{
    \begin{minipage}{\textwidth}
        \hspace{0.02\textwidth}
        \includegraphics[width=0.46\linewidth]{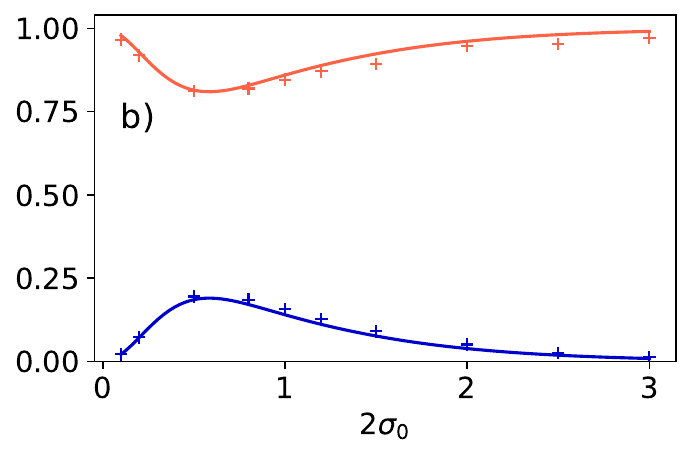}
        \hspace{0.04\textwidth}
        \includegraphics[width=0.46\linewidth]{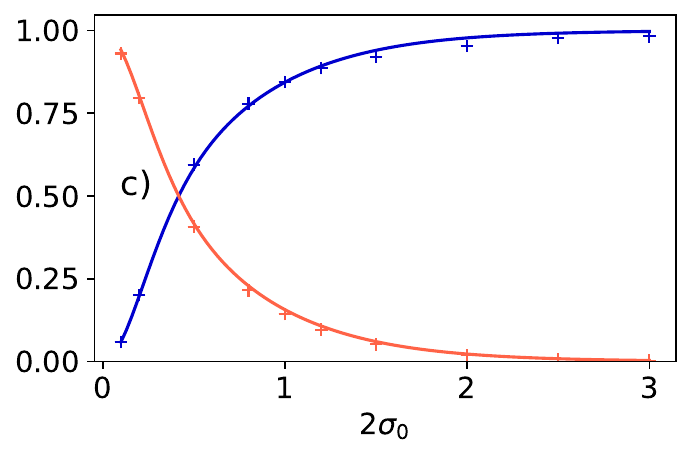}
    \end{minipage}}
    \caption{Predicted and measured kinetic energies in the case of a Gaussian shear layer, analogous to Figure \ref{fig:SingleLayer_Results}.}
\label{fig:MultiLayer_Results}
\end{figure}

In this section, we extend our analysis to a continuous geostrophic shear profile given by a Gaussian, see \eqref{eq:GaussShear}.
In figure \ref{fig:MultiLayer_Results} (upper left) we plot the predicted energy reflection coefficient using our analytic theory, and via DNS with Dedalus. 

\begin{figure}[ht!]
    \centerline{
    \includegraphics[width=0.5\linewidth]{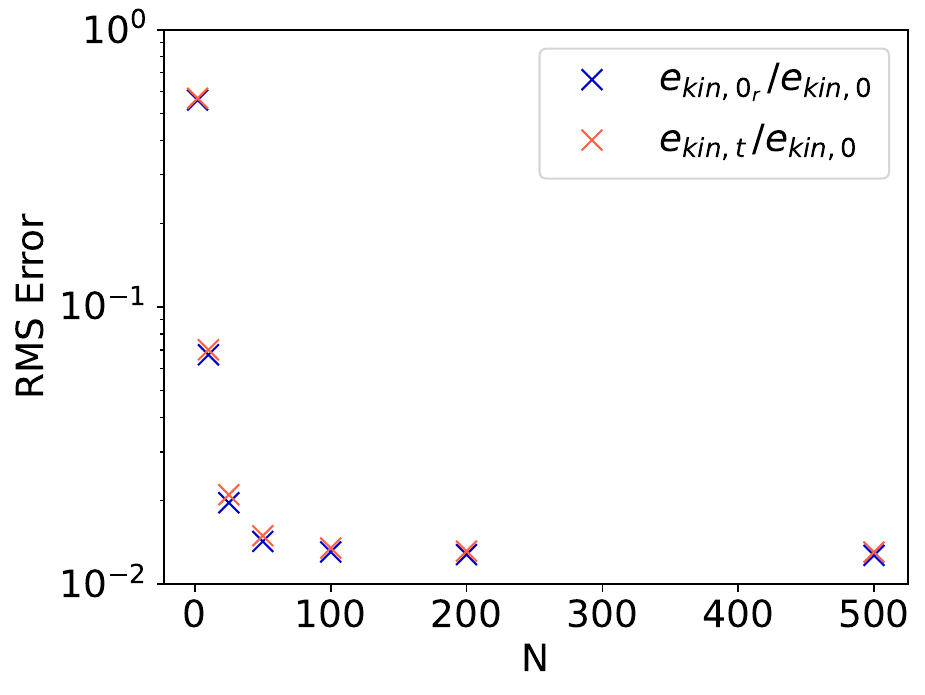}}
    \caption{Root-mean squared error between the measured and the theoretically predicted coefficients for the data presented in Figure \ref{fig:MultiLayer_Results} (a) using various refinements $N$ of the piecewise constant shear layer.}
\label{fig:RMS}
\end{figure}

We find good agreement between the simulations with continuous shear profile and the analytic model with $N=100$ interfaces (figure \ref{fig:MultiLayer_Results}, lower left and right), implying that our multi-layer approach is accurate. 
In figure \ref{fig:RMS} we compare the measured coefficients with our predictions for different numbers of interfaces $N$.
This shows the analytical theory gives an accurate prediction when $N>100$. The convergence with increasing $N$ is consistent with the results of \citet{belyaev2015properties} who studied an analogous problem for gravity waves. As for why the errors stagnate for $N\gtrsim 100$ and do not decrease further, we have no interpretation at this point. 

Similar to the single layer case, we find (1) significant reflected energy when $\delta \omega > \delta \omega_c$, and (2) transmission via tunnelling when $\delta \omega < \delta \omega_c$.
While weak shears ($|\delta \omega| < |\delta \omega_c|$) still backscatter waves, this effect is only significant if the width of the shear layer, is close to or smaller than the wavelength ($2\sigma_0 
\lesssim 1$). In contrast, for a single sharp shear layer, we found periodic behaviour of the reflection coefficient due to interference of the reflected waves at the two interfaces.  For the smooth shear layer, if $2\sigma_0 \lesssim 1$, the wave interacts with the shear layer as if it is sharp, producing similar behaviour as above. However, if the scale of the waves is small compared to the thickness of the shear layer, the wave is simply refracted through the smooth transition.\\
Furthermore, if the width of the jet $2\sigma$ exceeds the wavelength of the waves, we obtain the result of ray theory: total transmission for subcritical shear and total reflection for supercritical shear \citep{kunze1985near}.
However,
our investigation shows the behaviour of long wavelength waves is not accurately represented by ray theory.
Even for subcritical shear, there might be significant back-scattering of energy if the width of the layer is on the order of the wavelength. Consequently, special attention must be paid to the propagation and scattering of inertial waves of wavelengths approaching the length scale of the background flow. This might be of significance in spherical shells where a cylindrical Stewartson layer is formed at the inner core boundary with a similar length scale as viscous inertial waves \citep{stewartson1957almost, stewartson1966almost, kerswell1995internal}.\\
Regions of supercritical shear might not fully inhibit inertial waves from propagating as the waves can tunnel through the regions if the width of these is smaller than their wavelength--with almost total transmission for $2\sigma_0 \sim 0.1$.
This suggests a supercritical geostrophic shear acts as a low-pass filter for inertial waves, as we will demonstrate below.\\
Furthermore, understanding the interaction of the inertial wavefield with a simple continuous shear layer allows us to comprehend the interaction with a more complex profile of $\partial_x U_0(x)$, e.g., the shear profile of a geostrophic jet which transitions from positive to negative shear values (or vice versa). We can conceive this shear layer as a stack of two shear layers---one with $\partial_x U_0<0$ and one where $\partial_x U_0>0$---such that their transmission coefficients multiply.

\subsection{Geostrophic shear layers as low-pass filters for wave-beams}
\begin{figure}[ht!]
    \centerline{\includegraphics[width=\textwidth]{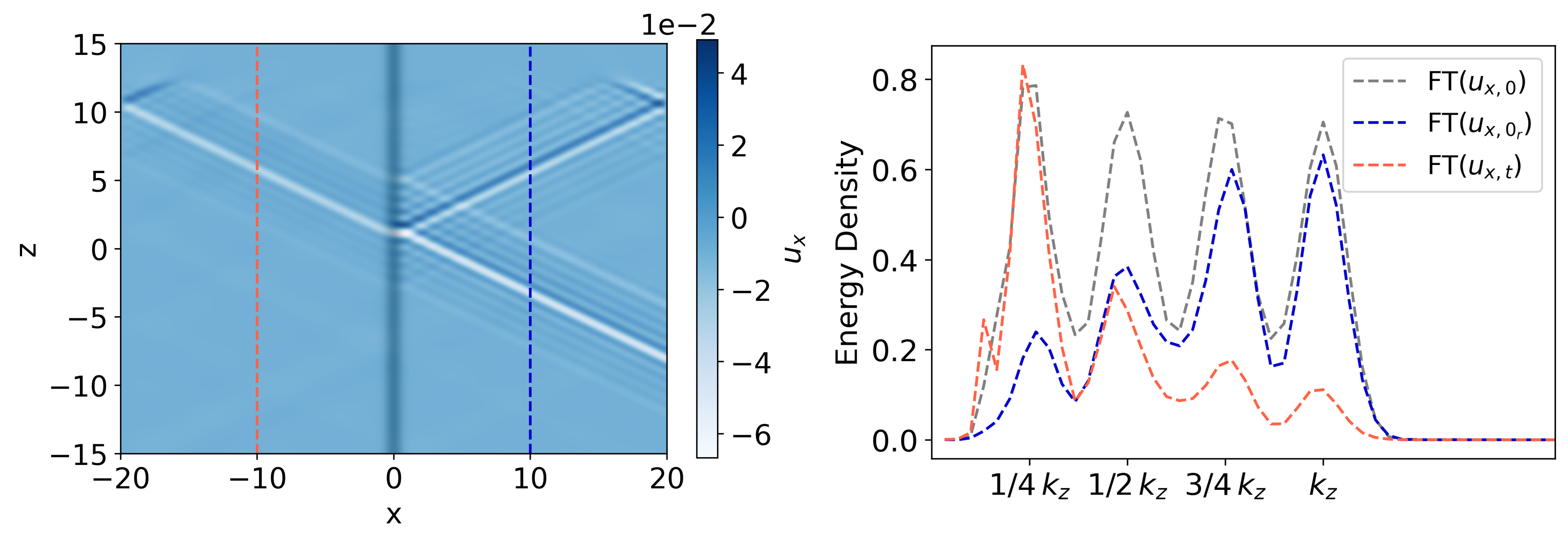}}
    \caption{Wavefield (left) and spectra (right) for the wave beam excited by the source function $F_4$. We take the FT along the two profiles at $x = \pm 10$.}
    \label{fig:Spectrum}
\end{figure}

To test the filtering ability of the Gaussian shear layer, we modify our source function $F(z,t)$ in \eqref{eq:Source} such that we now excite a wave beam including four distinct vertical wavenumbers. We excite the waves using
\begin{equation}
    u_x(x = 20,z,t) = F_4(z,t) = Ro\, \exp{\left(-\frac{(z-z_0)^2}{2d_\omega^2}\right)}\, \sum_{n = 1}^4 \sin(\frac{n}{4} k_z z - \omega t)
\end{equation}
Hence, the largest excited wavelength is $4\lambda$, i.e., spanning the width $2d_\omega$ of our source region. For the width of the jet, we choose $2\sigma_0 = 1$ and a supercritical shear value of $\delta \omega = -0.5$.\\
To measure the wavenumber content within the beams, we perform Fourier transforms (FTs) of $u_x$ along the two profiles at $|x| = 10$, to the right and left of the shear layer. To distinguish between the energies of the incident and reflected wave in the right profile, we set $u_x$ to zero outside the respective beam and then perform the FT. We display the profile lines in figure \ref{fig:Spectrum} (left), as well as the wavefield of $u_x$. In figure \ref{fig:Spectrum} (right), the FTs themselves can be seen.

The incident spectrum shows four prominent peaks from the four forced wavenumbers. As described above, the low wavenumber waves are preferentially transmitted through the jet, while the high wavenumbers are preferentially reflected. Hence, we found supercritical shear layers act as low-pass filters for wave beams.
This means that the inertial wave spectrum is sensitive to supercritical geostrophic shear layers with their thickness being an important parameter.

\section{Conclusion and Outlook} \label{sec:Conclusion}
In this paper, we describe how inertial waves interact with geostrophic shear layers. We first present an analytic calculation in which we approximate a continuous geostrophic shear layer by a series of discrete shear layers and validate these results with direct numerical simulations of waves interacting with continuous geostrophic shear.\\
We find two new phenomena, which have not been reported before in the context of inertial waves. 
\begin{enumerate}
\item \, For subcritical shear profiles ($\delta \omega > \delta \omega_c$), where ray theory predicts total transmission, we observe back-scattering of energy. This effect is strongest if the width of the shear layer is around one fourth of the horizontal wavelength.
That is, the wave reflects off both sides of the shear layer, leading to constructive interference.\\
\item \, For shear layers much thinner than the horizontal wavelength, the reflection coefficient diminishes and the waves tunnel through, even for supercritical shear ($\delta \omega < \delta \omega_c$). Evanescent waves transfer motion beyond the supercritical layer and form propagating inertial waves past it.
\end{enumerate}

\begin{figure}[ht!]
    \centerline{ 
    \includegraphics[width=0.3\linewidth]{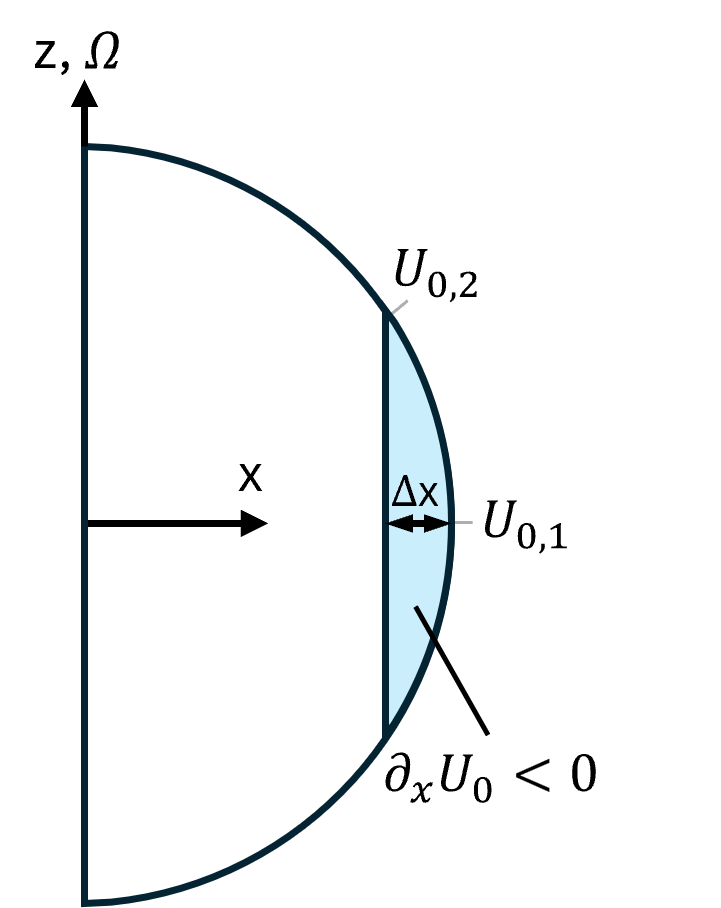}}
    \caption{Simplified geostrophic model of the shear layer introduced by Jupiter's equatorial jet}
    \label{fig:EquZone_Jupiter}
\end{figure}

\noindent These two processes cannot be described by ray theory, as the method is only valid for $\lambda$ smaller than the scale $L$ of the geostrophic shear \citep{kunze1985near}. Thus, we have both explored a new limit $\lambda \geq L$, while also reproducing the results of ray theory for $\lambda < L$.\\
Additionally, we demonstrate that thin, supercritical shear layers act as low-pass filters for inertial wave beams, allowing only low-wavenumber signals to pass. \\
One could raise the question of whether subsurface jets and currents in planetary fluid regions can be inferred from observing inertial wave spectra at the surface given the knowledge on the filtering nature of geostrophic currents.
As an example, let us estimate the importance of this filtering mechanism for Jupiter's equatorial jet.
From the data of \citet{tollefson2017changes} (PJ03), we obtain that the wind speed of Jupiter's equatorial jet increases from $U_{0,1} \approx 70\,\frac{\text{m}}{\text{s}}$ to $U_{0,2} \approx 110\,\frac{\text{m}}{\text{s}}$ between the equator and 6.5\,$^\circ$ north. Assuming that the wind patterns are geostrophic at first order and form cylindrical shear layers in Jupiter's interior (\citealt{kaspi2023observational}, see figure \ref{fig:EquZone_Jupiter}), this equatorial shear zone can be estimated to reach $\Delta x = 430$\,km deep into the interior, roughly the scale of wave-like features observed in Jupiter's atmosphere \citep{orton2020survey}.

\noindent We estimate the shear strength $\delta \omega_\text{Eq}$ dividing the velocity change $\Delta U_0 = (U_{0,1}-U_{0,2})$ by the width $\Delta x$ and obtain
\begin{equation}
    \frac{\delta \omega_\text{Eq}}{\Omega_\text{Jup}} = -1.05,
\end{equation}
where the characteristic frequency is $\Omega_\text{Jup} = 1.76\times10^{-4}\,\frac{1}{s}$, the rotation rate of Jupiter.\\
Hence, $\delta \omega_\text{Eq}$ is supercritical, independent of the frequency of the wave and we are in the regime of $\lambda \gtrsim \Delta x$ in which ray-theory is not valid and wavenumber dependent filtering for pure inertial waves appears.\\
However, whether the discussed filtering mechanism significantly enters the dynamics and the observed waves are sensitive to the subsurface shear has to be discussed in a wider perspective: As the wave-like features on Jupiter are likely to be gravity-inertia waves, stratification and buoyancy forces are non-negligible and have to be considered in any analysis or inversion. Furthermore, most waves seem to travel in zonal direction and along-wind, introducing advection as an additional effect, which has not been considered in our work, using $k_y = 0$ (both effects of advection and stratification are, e.g., included by \citet{simon2018new} to model gravity-inertia waves on Jupiter, but not the effect of flow shear).\\
Consequently, further research should aim at bringing together all of the building blocks describing the interaction of internal waves with anomalous layers in the scale of their wavelength, including the effect of stratification and mixed layers \citep{sutherland2004internal, belyaev2015properties}, along-flow propagation (e.g. \citet{simon2018new}), and background shear, the latter of which has been established in this article. 


\section*{Funding statement}
This work was supported by the Simons Foundation (grant SFI-MPS-T-MPS-00007353).
LK is supported by the Swiss National Science Foundation (SNF grant 200021-228104/ 10001286).
DL is partially supported by the National Science Foundation AAG (grant AST-2405812) and the Sloan Foundation (grant FG-2024-21548).
Computational cost were supported by ETH Zurich (grant ETH-04 22-1).

\section*{Competing interests}
The authors declare no conflict of interests.

\section*{Data availability statement}
To receive any computational notebooks or data files related to the results of this article, please contact the corresponding author.

\section*{Author ORCID}
Lennart Kira, \url{https://orcid.org/0009-0007-3120-6529}\\
Jerome Noir, \url{https://orcid.org/0000-0001-9977-0360}\\
Daniel Lecoanet, \url{https://orcid.org/0000-0002-7635-9728}



\begin{thebibliography}{}


\bibitem[Baruteau and Rieutord (2012)]{baruteau2012inertial}
{\sc Baruteau, C. and Rieutord, M.} 2017 {Inertial waves in a differentially rotating spherical shell}, {\it arXiv preprint arXiv:1203.4347} \url{https://doi.org/10.48550/arXiv.1203.4347}

\bibitem[Belyaev et al. (2015)]{belyaev2015properties}
{\sc Belyaev, M. A., Quataert, E., Fuller, J.} 2015 {The properties of g-modes in layered semiconvection}, {\it MNARS}, {\bf 452}, pp. 2700-2711. \url{https://doi.org/10.1093/mnras/stv1446}

\bibitem[Bracamontes-Ramirez and Sutherland (2024)]{Bracamontes-Ramirez2024}
{\sc Bracamontes-Ramirez, J. and Sutherland, B. R.} 2024 {Transient internal wave excitation of resonant modes in a density staircase}, {\it Phys. Rev. Fluids} {\bf 9}, p. 064801 \url{https://doi.org/10.1103/PhysRevFluids.9.064801}

\bibitem[Burns et al. (2020)]{Dedalus}
{\sc Burns, K. J., Vasil, G. M., Oishi, J. S., Lecoanet, D. \& Brown, B. P.} 2020 {Dedalus: A flexible framework for numerical simulations with spectral methods}, {\it Phys. Rev. Res.}, {\bf 2} (2), 023068. \url{https://doi.org/10.1103/PhysRevResearch.2.023068}

\bibitem[Davidson (2013)]{davidson2013turbulence}
{\sc Davidson, P. A.} 2013 {Turbulence in Rotating, Stratified and Electrically Conducting Fluids} {\it Cambridge University Press}


\bibitem[Gonella (1972)]{gonella1972rotary}
{\sc Gonella, J.} 1972 {A rotary-component method for analysing meteorological and oceanographic vector time series}, {\it Deep-Sea Res. Oceanogr. Abstr.}, {\bf 19} (12), pp. 833--846. \url{https://doi.org/10.1016/0011-7471(72)90002-2}

\bibitem[Hanson et al. (2022)]{hanson2022discovery}
{\sc Hanson, C. S., Hanasoge, S. \& Sreenivasan, K. R.} 2022 {Discovery of high-frequency retrograde vorticity waves in the Sun}, {\it Nat. Astron.}, {\bf 6} (6), pp. 708--714. \url{https://doi.org/10.1038/s41550-022-01632-z}

\bibitem[Kaspi et al. (2023)]{kaspi2023observational}
{\sc Kaspi, Y., Galanti, E., Park, R. S., Duer, K., Gavriel, N., Durante, D., Iess, L., Parisi, M., Buccino, D. R., Guillot, T. and others} 2023 {Observational evidence for cylindrically oriented zonal flows on Jupiter}, {\it Nat. Astron.}, {\bf 7} (12), pp. 1463-1472. \url{https://doi.org/10.1038/s41550-023-02077-8}

\bibitem[Kerswell (1995)]{kerswell1995internal}
{\sc Kerswell, R. R.} 1995 {On the internal shear layers spawned by the critical regions in oscillatory Ekman boundary layers}, {\it J. Fluid Mech.}, {\bf 298}, pp. 311--325. \url{https://doi.org/10.1017/S0022112095003326}

\bibitem[Kunze (1985)]{kunze1985near}
{\sc Kunze, E.} 1985 {Near-inertial wave propagation in geostrophic shear}, {\it J. Phys. Oceanogr.}, {\bf 15} (5), pp. 544--565. \url{https://doi.org/10.1175/1520-0485(1985)015%3C0544:NIWPIG%3E2.0.CO;2}

\bibitem[Le Bars et al. (2015)]{le2015flows}
{\sc Le Bars, M., Cébron, D. \& Le Gal, P.} 2015 {Flows driven by libration, precession, and tides}, {\it Annu. Rev. Fluid Mech.}, {\bf 47} (1), pp. 163--193. \url{https://doi.org/10.1146/annurev-fluid-010814-014556}

\bibitem[Le Bars \& Lecoanet (2020)]{le2020fluid}
{\sc Le Bars, M. \& Lecoanet, D.} 2020 {Fluid Mechanics of Planets and Stars} {\it Springer}

\bibitem[Munk (1981)]{munk1981internal}
{\sc Munk, W. H.} 1981 {Internal waves and small-scale processes}, {\it Evolution of Physical Oceanography, MIT Press.}

\bibitem[Nosan et al. (2021)]{nosan2021evanescent}
{\sc Nosan, {\v{Z}}., Burmann, F., Davidson, P. A. \& Noir, J.} 2021 {Evanescent inertial waves}, {\it J. Fluid Mech.}, {\bf 918}, R2. \url{https://doi.org/10.1017/jfm.2021.343}

\bibitem[Olbers (1981)]{olbers1981formal}
{\sc Olbers, D. J.} 1981 {A formal theory of internal wave scattering with applications to ocean fronts}, {\it J. Phys. Oceanogr.}, {\bf 11} (8), pp. 1078--1099. \url{https://doi.org/10.1175/1520-0485(1981)011%3C1078:AFTOIW%3E2.0.CO;2}

\bibitem[Orton et al. (2020)]{orton2020survey}
{\sc Orton, G. S., Tabataba-Vakili, F., Eichstädt, G., Rogers, J., Hansen, C. J., Momary, T. W., Ingersoll, A. P., Brueshaber, S., Wong, M. H., Simon, A. A. {\it et al.}} 2020 {A survey of small-scale waves and wave-like phenomena in Jupiter's atmosphere detected by JunoCam}, {\it J. Geophys. Res. Planets}, {\bf 125} (7), e2019JE006369. \url{https://doi.org/10.1029/2019JE006369}

\bibitem[Ouazzani et al. (2020)]{Ouazzani2020}
{\sc Ouazzani, R. M., Ligni{\`e}res, F., Dupret, M. A., Salmon, S.~J.~A.~J., Ballot, J., Christophe, S., Takata, M.} 2020 {First evidence of inertial modes in {\ensuremath{\gamma}} Doradus stars: The core rotation revealed}, {it A\&A} {\bf 640}, p. A49 \url{https://doi.org/10.1051/0004-6361/201936653}

\bibitem[Phillips (1963)]{phillips1963energy}
{\sc Phillips, O. M.} 1963 {Energy transfer in rotating fluids by reflection of inertial waves}, {\it Phys. Fluids}, {\bf 6} (4), pp. 513--520. \url{https://doi.org/10.1063/1.1706766}

\bibitem[Rogers et al. (2016)]{rogers2016dispersive}
{\sc Rogers, J. H., Fletcher, L. N., Adamoli, G., Jacquesson, M., Vedovato, M. \& Orton, G. S.} 2016 {A dispersive wave pattern on Jupiter’s fastest retrograde jet at 20°S}, {\it Icarus}, {\bf 277}, pp. 354--369. \url{https://doi.org/10.1016/j.icarus.2016.05.028}

\bibitem[Saio et al. (2021)]{Saio2021}
{\sc Saio, H., Takata, M., Lee, U., Li, G., Van Reeth, T.} 2021 {Rotation of the convective core in {\ensuremath{\gamma}} Dor stars measured by dips in period spacings of g modes coupled with inertial modes}, {\it MNRAS} {\bf 502} (4), pp. 5856-5874 

\bibitem[Simon et al. (2018)]{simon2018new}
{\sc Simon, A. A., Hueso, R., Inurrigarro, P., Sanchez-Lavega, A., Morales-Juberias, R., Cosentino, R., Fletcher, L. N., Wong, M. H., Hsu, A. I., De Pater, I. others} 2018 {A new, long-lived, Jupiter mesoscale wave observed at visible wavelengths}, {\it AJ}, {\bf 156} (2), p. 79. \url{https://doi.org/10.3847/1538-3881/aacaf5}

\bibitem[Stewartson (1957)]{stewartson1957almost}
{\sc Stewartson, K.} 1957 {On almost rigid rotations}, {\it J. Fluid Mech.}, {\bf 3} (1), pp. 17--26. \url{https://doi.org/10.1017/S0022112057000452}

\bibitem[Stewartson (1966)]{stewartson1966almost}
{\sc Stewartson, K.} 1966 {On almost rigid rotations. Part 2}, {\it J. Fluid Mech.}, {\bf 26} (1), pp. 131--144. \url{https://doi.org/10.1017/S0022112066001137}

\bibitem[Sutherland and Yewchuk (2004)]{sutherland2004internal}
{\sc Sutherland, B. R., Yewchuk, K.} 2004 { Internal wave tunnelling}, {\it J. of Fluid Mech.}, {\bf 511}, pp. 125-134. \url{https://doi.org/10.1017/S0022112004009863}

\bibitem[Sutherland (2016)]{Sutherland2016}
{\sc Sutherland, B. R.} 2016 {Internal wave transmission through a thermohaline staircase}, {\it Phys. Rev. Fluids} {\bf 1}, p. 013701 \url{https://doi.org/10.1103/PhysRevFluids.1.013701}

\bibitem[Tollefson et al. (2017)]{tollefson2017changes}
{\sc Tollefson, J., Wong, M. H., de Pater, I., Simon, A. A., Orton, G. S., Rogers, J. H., Atreya, S. K., Cosentino, R. G., Januszewski, W., Morales-Juberias, R. and others} 2017 {Changes in Jupiter's zonal wind profile preceding and during the Juno mission}, {\it Icarus}, {\bf 296}, pp. 163-178. \url{https://doi.org/10.1016/j.icarus.2017.06.007}

\bibitem[Triana et al. (2022)]{triana2022identification}
{\sc Triana, S. A., Guerrero, G., Barik, A. \& Rekier, J.} 2022 {Identification of inertial modes in the solar convection zone}, {\it Astrophys. J. Lett.}, {\bf 934} (1), L4. \url{https://doi.org/10.3847/2041-8213/ac7dac}

\bibitem[Young \& Jelloul (1997)]{young1997propagation}
{\sc Young, W. R. \& Jelloul, M. B.} 1997 {Propagation of near-inertial oscillations through a geostrophic flow}, {\it J. Mar. Res.}, {\bf 55} (4). \url{https://elischolar.library.yale.edu/journal_of_marine_research/2242}\url{https://doi.org/10.1093/mnras/stab482}

\end{thebibliography}
\end{document}